
\input harvmac
\input epsf
\input amssym
%
%
\noblackbox
\newcount\figno
\figno=0
\def\fig#1#2#3{
\par\begingroup\parindent=0pt\leftskip=1cm\rightskip=1cm\parindent=0pt
\baselineskip=11pt
\global\advance\figno by 1
\midinsert
\epsfxsize=#3
\centerline{\epsfbox{#2}}
\vskip -21pt
{\bf Fig.\ \the\figno: } #1\par
\endinsert\endgroup\par
}
\def\figlabel#1{\xdef#1{\the\figno}}
\def\encadremath#1{\vbox{\hrule\hbox{\vrule\kern8pt\vbox{\kern8pt
\hbox{$\displaystyle #1$}\kern8pt}
\kern8pt\vrule}\hrule}}

\def\frac#1#2{{#1 \over #2}}

\def\p{\partial}
\def\semi{\subset\kern-1em\times\;}

\def\sqr#1#2{{\vcenter{\vbox{\hrule height.#2pt
\hbox{\vrule width.#2pt height#1pt \kern#1pt \vrule width.#2pt}
\hrule height.#2pt}}}}

\def\p{\partial}

\def\rh{\hat{r}}

\def\p{\partial}

\def\ls{\ell_{S^p} }

\def\eps{\epsilon}
\def\wb{\overline{w}}

%

%
\def\ZZ{\Bbb{Z}}

\def\zb{\overline{z}}


\lref\FreedTG{
  D.~Freed, J.~A.~Harvey, R.~Minasian and G.~W.~Moore,
  ``Gravitational anomaly cancellation for M-theory fivebranes,''
  Adv.\ Theor.\ Math.\ Phys.\  {\bf 2}, 601 (1998)
  [arXiv:hep-th/9803205].
}

\lref\SkenderisWP{
  K.~Skenderis,
  ``Lecture notes on holographic renormalization,''
  Class.\ Quant.\ Grav.\  {\bf 19}, 5849 (2002)
  [arXiv:hep-th/0209067].
}

\lref\HarveyBX{
  J.~A.~Harvey, R.~Minasian and G.~W.~Moore,
  ``Non-abelian tensor-multiplet anomalies,''
  JHEP {\bf 9809}, 004 (1998)
  [arXiv:hep-th/9808060].
}

\lref\MaldacenaBW{
  J.~M.~Maldacena and A.~Strominger,
  ``AdS(3) black holes and a stringy exclusion principle,''
  JHEP {\bf 9812}, 005 (1998)
  [arXiv:hep-th/9804085].
}

\lref\KrausDI{
  P.~Kraus, F.~Larsen and R.~Siebelink,
  ``The gravitational action in asymptotically AdS and flat spacetimes,''
  Nucl.\ Phys.\ B {\bf 563}, 259 (1999)
  [arXiv:hep-th/9906127].
}

\lref\BalasubramanianRE{
  V.~Balasubramanian and P.~Kraus,
  ``A stress tensor for anti-de Sitter gravity,''
  Commun.\ Math.\ Phys.\  {\bf 208}, 413 (1999)
  [arXiv:hep-th/9902121].
}

\lref\BalasubramanianRT{
  V.~Balasubramanian, J.~de Boer, E.~Keski-Vakkuri and S.~F.~Ross,
  ``Supersymmetric conical defects: Towards a string theoretic description  of
  black hole formation,''
  Phys.\ Rev.\ D {\bf 64}, 064011 (2001)
  [arXiv:hep-th/0011217].
}

\lref\MaldacenaDR{
  J.~M.~Maldacena and L.~Maoz,
  ``De-singularization by rotation,''
  JHEP {\bf 0212}, 055 (2002)
  [arXiv:hep-th/0012025].
}

\lref\CveticDM{
  M.~Cvetic, H.~Lu and C.~N.~Pope,
  ``Consistent Kaluza-Klein sphere reductions,''
  Phys.\ Rev.\ D {\bf 62}, 064028 (2000)
  [arXiv:hep-th/0003286].
}

\lref\Bott{R.~Bott and A.S.~Cattaneo, ``Integral invariant of
3-manifolds," [arxiv:dg-ga/9710001].}

\lref\CveticXH{
  M.~Cvetic and F.~Larsen,
  ``Near horizon geometry of rotating black holes in five dimensions,''
  Nucl.\ Phys.\ B {\bf 531}, 239 (1998)
  [arXiv:hep-th/9805097].
}

\lref\BalasubramanianEE{
  V.~Balasubramanian and F.~Larsen,
  ``Near horizon geometry and black holes in four dimensions,''
  Nucl.\ Phys.\ B {\bf 528}, 229 (1998)
  [arXiv:hep-th/9802198].
}

\lref\HenningsonGX{
  M.~Henningson and K.~Skenderis,
  ``The holographic Weyl anomaly,''
  JHEP {\bf 9807}, 023 (1998)
  [arXiv:hep-th/9806087].
}

\lref\Elitzur{
  S.~Elitzur, G.~W.~Moore, A.~Schwimmer and N.~Seiberg,
  ``Remarks On The Canonical Quantization Of The Chern-Simons-Witten Theory,''
  Nucl.\ Phys.\ B {\bf 326}, 108 (1989).}

\lref\NastaseCB{
  H.~Nastase, D.~Vaman and P.~van Nieuwenhuizen,
  ``Consistent nonlinear K K reduction of 11d supergravity on AdS(7) x S(4)
  and self-duality in odd dimensions,''
  Phys.\ Lett.\ B {\bf 469}, 96 (1999)
  [arXiv:hep-th/9905075].
}

\lref\HollandsWT{
  S.~Hollands, A.~Ishibashi and D.~Marolf,
  ``Comparison between various notions of conserved charges in  asymptotically
  AdS-spacetimes,''
  Class.\ Quant.\ Grav.\  {\bf 22}, 2881 (2005)
  [arXiv:hep-th/0503045].
}

\lref\PapadimitriouII{
  I.~Papadimitriou and K.~Skenderis,
  ``Thermodynamics of asymptotically locally AdS spacetimes,''
  JHEP {\bf 0508}, 004 (2005)
  [arXiv:hep-th/0505190].
}
\lref\TaylorRobinsonFE{
  M.~Taylor-Robinson,
  ``Higher-dimensional formulation of counterterms,''
  arXiv:hep-th/0110142.
}

\lref\SkenderisUY{
  K.~Skenderis and M.~Taylor,
  ``Kaluza-Klein holography,''
  JHEP {\bf 0605}, 057 (2006)
  [arXiv:hep-th/0603016].
}

\lref\SkenderisDI{
  K.~Skenderis and M.~Taylor,
  ``Holographic Coulomb branch vevs,''
  arXiv:hep-th/0604169.
}

\lref\KL{P.~Kraus and F.~Larsen, to appear.}

\lref\KrausZM{
  P.~Kraus and F.~Larsen,
  ``Holographic gravitational anomalies,''
  JHEP {\bf 0601}, 022 (2006)
  [arXiv:hep-th/0508218].
}

\lref\KrausVZ{
  P.~Kraus and F.~Larsen,
  ``Microscopic black hole entropy in theories with higher derivatives,''
  JHEP {\bf 0509}, 034 (2005)
  [arXiv:hep-th/0506176].
}

\lref\BanadosFE{
  M.~Banados, O.~Miskovic and S.~Theisen,
   ``Holographic currents in first order gravity and finite Fefferman-Graham
  expansions,''
  JHEP {\bf 0606}, 025 (2006)
  [arXiv:hep-th/0604148].
}

\lref\deWitIY{
  B.~de Wit and H.~Nicolai,
  ``THE CONSISTENCY OF THE S**7 TRUNCATION IN d = 11 SUPERGRAVITY,''
  Nucl.\ Phys.\ B {\bf 281}, 211 (1987).
}



\Title{\vbox{\baselineskip12pt
}} {\vbox{\centerline {Generating Charge from Diffeomorphisms}}} \centerline{James
Hansen\foot{jhansen@physics.ucla.edu} and Per
Kraus\foot{pkraus@physics.ucla.edu}}

\bigskip
\centerline{${}^1$\it{Department of Physics and Astronomy,
UCLA,}}\centerline{\it{ Los Angeles, CA 90095-1547, USA.}}

\baselineskip15pt

\vskip .3in

\centerline{\bf Abstract}

We unravel some subtleties involving the definition of sphere
angular momentum charges in AdS$_q\times S^p$ spacetimes, or
equivalently, R-symmetry charges in the dual boundary CFT. In the
AdS$_3$ context, it is known that charges can be generated by
coordinate transformations, even though the underlying theory is
diffeomorphism invariant.  This is the bulk version of spectral
flow in the boundary CFT.  We trace this behavior back to special
properties of the p-form field strength supporting the solution,
and derive the explicit formulas for angular momentum charges.
This analysis also reveals  the higher
dimensional origin of three dimensional Chern-Simons terms and
of chiral anomalies in the boundary theory.

\baselineskip14pt

\newsec{Introduction}

AdS$_q \times S^p$ spacetimes in string/M-theory arise as solutions
of gravity coupled to a p-form field strength, as described by the Euclidean
signature action\foot{In odd dimensions a Chern-Simons term will also
play an important role (see below).}
\eqn\za{ S = {1 \over 16\pi G_{q+p}} \int\!d^{q+p}x \left
(\sqrt{g}R+ \half{\star} G_p \wedge G_p \right)~.}
 This
theory admits Freund-Rubin type solutions:
\eqn\zb{ \eqalign{ ds^2 &= ds_{AdS_q}^2 + ds_{S^p}^2 \cr G_p &= Q
\epsilon_{S^p}~.}}
In this paper we are primarily interested in the case of
asymptotically, locally, AdS$_3\times S^p$ spacetimes.  Such
geometries, with suitable boundary conditions, have a local $SO(p+1)$
group of symmetries associated with isometries of the p-sphere,
and corresponding conserved charges.  Here we aim to give an
explicit expression for these conserved charges.

To appreciate that this problem is more subtle than one might
guess, observe the following.  Start with the solution \zb, which
has vanishing $SO(p+1)$ charges.  Now perform a simple coordinate
transformation that mixes up the sphere and AdS coordinates. Since
we are working in the context of a diffeomorphism invariant
theory, it seems natural to expect that the $SO(p+1)$ charges will
continue to vanish after the coordinate transformation.  But this
expectation clashes with the charges usually assigned to standard
solutions of this form,\foot{These charge assignments are
typically made by comparing with the angular momenta of
asymptotically flat solutions with a given near horizon geometry.
Since angular momenta are quantized they are expected to be
unchanged upon taking the near horizon limit. In this work we make
no reference to auxiliary asymptotically flat solutions.} and with
basic aspects of the AdS$_3$/CFT$_2$ duality. For example,
rotating  BPS black hole solutions in the D1-D5 system look
locally like a coordinate transformation of AdS$_3\times S^3$, yet
carry nonzero charge \CveticXH. In the boundary CFT description
there is the phenomenon of ``spectral flow", which is a
relabelling of states and symmetry generators that shifts the
R-charges.      The gravitational description of spectral flow is
known to be a coordinate transformation of the sort we have just
described \refs{\BalasubramanianRT,\MaldacenaDR}.

The resolution of this puzzle, and the route to obtaining
acceptable formulas for conserved charges, involves several
ingredients. The solutions described by  \za-\zb\  will in fact carry
zero charge after a coordinate transformation ---
extra structure is required to induce the charges. Our two basic examples are
AdS$_3\times S^3$ and  AdS$_3\times S^2$. In the $S^3$ case it is
crucial that $G_3$ also have flux on AdS$_3$, while for $S^2$ we
need to include in the action a 5-dimensional Chern-Simons term,
$\int C_1 \wedge G_2 \wedge G_2$. In both cases the crux of our
analysis is then  a careful treatment of the p-form field
strength.  To obtain a satisfactory gauge invariant theory on
AdS$_3$ after reduction on the sphere, we are forced to have $C_{p-1}$ and/or
$G_p$
transform in a nontrivial way under $SO(p+1)$.
 In the $S^2$ case $G_2$ will be $SO(3)$ invariant, but
$C_1$ will not be, and the presence of the 5-dimensional
Chern-Simons term then induces the nonzero charge.  In the $S^3$
case the charge will arise from the $SO(4)$ noninvariance of
$G_3$.   Our treatment  of these two cases will admit a
generalization to AdS$_{2n-1} \times S^{2n-1}$ and
AdS$_{4n-1}\times S^{2n}$.

The basic tools for obtaining gauge invariant actions have been
developed in the context of M5-brane anomaly  cancellation
\refs{\FreedTG,\HarveyBX} and  consistent Kaluza-Klein sphere
reductions (e.g. \refs{\deWitIY,\NastaseCB,\CveticDM}), and we
will adapt them for our purposes.  The
 Kaluza-Klein procedure
produces gauged supergravity theories, which contain three
dimensional Chern-Simons terms for the $SO(p+1)$ gauge fields.  As
we discuss in the next section, the construction of charges with
the desired properties follows quite straightforwardly from this
three-dimensional perspective.  In particular, the spectral flow
behavior is linked to the fact that Chern-Simons terms are only
gauge invariant up to boundary terms. The basic challenge for us
will be to obtain these results directly from the higher
dimensional setup without performing a Kaluza-Klein reduction. To
do  so we can use previous insights \HarveyBX\ on the higher
dimensional origin of Chern-Simons terms.   We should emphasize
that although we will be using some of the methods developed in
the context of consistent Kaluza-Klein sphere reductions, our
conclusions will be more general.  In particular, since we will
only need to make reference to the asymptotic behavior of the
fields, we can allow for the presence of additional fields beyond
those appearing in a consistent truncation ansatz, as long as they
take fixed values at infinity.   This is fortunate, since in some
of the cases we discuss no complete consistent truncation has so
far been derived in the literature.

To forestall a possible confusion, we remark that there is another
unrelated context in which charges can be induced by coordinate
transformations.  Gravity in odd-dimensional AdS spacetimes has a
conformal anomaly arising from the need to regulate and subtract
large volume divergences in the action \HenningsonGX.   The result is  that the
gravitational action is not invariant under all coordinate
transformations, specifically those that act as a Weyl
transformation of the conformal boundary metric. It is therefore
not too surprising that coordinate transformations can shift the
charges associated with AdS energy and angular momentum, in
agreement with the expected anomalous transformation law of the
stress tensor \BalasubramanianRE.  Our case is different in that our coordinate
transformations will not act as Weyl transformation on the
boundary, and so  potential violations of diffeomorphism
invariance will play no role.

\newsec{Currents and charges in AdS$_3$ gravity}

We begin with  a discussion of currents and charges in the
effective three dimensional description of an underlying higher
dimensional theory.\foot{This is based on the more complete
analysis in \KL.   Note also that an analogous treatment of {\it
nonchiral} currents appears in \BanadosFE. Here we are concerned
with chiral currents, since these appear in the relevant AdS/CFT
examples.   } We work in the framework of holographic
renormalization \refs{ \HenningsonGX,\BalasubramanianRE} (for a
review see \SkenderisWP, and for additional work on definining
conserved charges see \refs{\HollandsWT,\PapadimitriouII} ). Our
goal will then be to reproduce these results from the higher
dimensional perspective.

The relevant terms in the action for the metric and 1-form
potential are
\eqn\aa{ S = {1 \over 16\pi G_3} \int\! d^3x\sqrt{g} \left(R-{2\over
\ell^2}\right) -{ik \over 4\pi} \int\! d^3 x~ \Tr(AdA+{2 \over 3
}A^3 ) + \ldots + S_{{\rm bndy}}~.}
We are working in Euclidean signature.   The $\ldots$ terms refer to
 contributions from other matter fields and possible higher
derivative terms  that will not contribute to our discussion of
charges, since these are controlled by the leading long distance
part of the Lagrangian.  The need for various boundary terms is also
indicated, and will be discussed in more detail below.

The Chern-Simons term is defined with respect to an $SU(2)$ gauge
group, which either can be thought of as the isometry group of an
$S^{2}$, or as one factor in the $SO(4)\approx SU(2)\times SU(2)$
isometry group of an $S^3$.    Invariance of the action  under
large gauge transformations requires that $k$ be an integer, which
we will take to be positive. The gauge field components are given
by
 $A=A^a{i\over 2}\sigma^a$.

The metric is taken to be asymptotically AdS$_3$ in the sense that
it takes the Fefferman-Graham form
\eqn\ab{ ds^2 = d\eta^2 +e^{2\eta/\ell} g^{(0)}_{\alpha\beta}
dx^\alpha dx^\beta + g^{(2)}_{\alpha\beta} dx^\alpha dx^\beta +
\ldots~.}
The gauge fields admit the expansion
\eqn\ac{ A = A^{(0)} + e^{-2\eta/\ell} A^{(2)} +\ldots~,}
and we choose the gauge $A_\eta =0$.\foot{Choosing a gauge is not
quite as innocuous as it sounds, since this theory is anomalous under
 gauge transformation that are nonzero at the boundary.
   It is perhaps better to say that we are  deciding  to look
just at solutions of this form.}

   Analysis of the field
equations (including the effect of Maxwell type terms) shows  that
$A^{(0)}$ is a flat connection; that is, the field strength
corresponding to \ac\ falls off as $e^{-2\eta/\ell}$.  This falloff
of the field strength implies that Maxwell and higher derivative terms
in the action will give no contributions to the on-shell variation of the
action, since the relevant surface integrals vanish.  So the analysis
that follows holds in complete generality.

We  define a stress tensor and current by evaluating the on-shell
variation of the action. When the equations of motion are satisfied,
the variation takes the form
\eqn\ae{ \delta S =  \int_{\p AdS}\! d^2x \sqrt{g^{(0)}}\left({1
\over 2} T^{\alpha\beta} \delta g^{(0)}_{\alpha\beta}  +{i\over
2\pi} J^{\alpha a} \delta A^{(0)a}_\alpha \right)~.}
Indices are raised and lowered with the conformal boundary metric
$g^{(0)}_{\alpha\beta}$. To put the variation in the above form we
need to add appropriate boundary terms to the action, as was
indicated in \aa.   As is well known \refs{
\HenningsonGX,\BalasubramanianRE}, the gravitational boundary term
is
\eqn\af{S_{\rm bndy}^{\rm grav} = {1 \over 8\pi G_3} \int_{\p AdS}\!
d^2x \sqrt{g} \left( \Tr K -{1 \over \ell}\right)~,}
where $K$ is the extrinsic curvature of the boundary.

For reasons that we will explain momentarily, it is also natural to
include the boundary term
\eqn\ag{ S_{\rm bndy}^{\rm gauge} =- {k \over 16\pi} \int_{\p AdS} \!
d^2 x \sqrt{g} g^{\alpha\beta} A^{a}_\alpha A^{a}_\beta~.}

With these boundary terms, the on-shell variation of the action yields
\eqn\ah{\eqalign{  T_{\alpha\beta} &= {1 \over 8\pi G\ell} \left(
g^{(2)}_{\alpha\beta}- \Tr (g^{(2)}) g_{\alpha\beta}^{(0)}\right)+
{k \over 8\pi}(A^{(0)a}_\alpha A^{(0)a}_\beta -\half A^{(0)a
\gamma}A^{(0)a}_\gamma g^{(0)}_{\alpha\beta}) \cr J^a_\alpha & =
{ik\over 4} (A^{(0) a}_\alpha -i \epsilon_\alpha^{~\beta}
A^{(0)a}_\beta )~.
 }}
To appreciate the role of the boundary term \ag, work in conformal
gauge: $g^{(0)}_{\alpha\beta}dx^\alpha dx^\beta = dw d\wb$.  Then
the current is
\eqn\ai{ J^a_w = {ik \over 2} A^{(0)a}_w~,\quad J^a_{\wb} =0~.}
The coefficient in front of \ag\ was chosen to set to zero the
anti-holomorphic component of the current. This is desirable from
several points of view. First, it is a standard fact from the
quantization of  Chern-Simons theory (e.g. \Elitzur) that $A_w$ and
$A_{\wb}$ are canonically conjugate in the sense that  the variation
of the action takes the form $\delta S\sim  p\delta q \sim A_w
\delta A_{\wb}$, which is consistent with \ai. This means that we
can adopt a variational principle in which the boundary conditions
are set by $A_{\wb}$ only. Fixing boundary conditions for both
components of $A$ is problematic in that there will typically not be
any smooth solution consistent with the chosen boundary conditions.
   Second, in the
context of CFTs dual to the bulk AdS$_3$ theory, the level $k$
$SU(2)$ current algebra is indeed holomorphic. In this regard, we
also note that \ai\ gives the correct chiral anomaly
\eqn\aj{ D_{\wb} J^a_w= {ik \over 2} \p_w A^{(0)a}_{\wb}~,}
where we used $F^{(0)} =0$.

Given the current we can define a charge.  In conformal gauge the charge is simply
\eqn\ak{ J^a_0 = \oint {dw \over 2\pi i} J^a_w = {ik\over 2}  \oint
{dw \over 2\pi i} A^{(0)a}_w~,}
where the contour goes around the AdS$_3$ boundary cylinder. The
charge is therefore equivalent to the gauge holonomy. From \aj\ we
see that the charge is conserved if $ A^{(0)a}_{\wb}=0$. The charges
obey the $SU(2)$ Lie algebra\foot{The factor of $i/2\pi$ is \ae\ was
chosen to bring the algebra to the standard form.}
\eqn\al{ [J^a_0,J^b_0] =i \epsilon^{abc} J^c_0~.}
More generally the modes $J_n^a = \oint {dw \over 2\pi i} w^n
J^a_w$ obey an $SU(2)$ current algebra at level $k$:
\eqn\ala{ [J^a_n,J^b_m] =\half m k \delta_{m,-n}\delta^{ab}+ i
\epsilon^{abc} J^c_{n+m}~.}

\ah\ shows that the gauge field  contributes to the stress tensor as
\eqn\al{ T_{ww} = {k \over 8\pi} A^a_w A^a_{w}~, \quad T_{\wb \wb}
= {k \over 8\pi} A^a_{\wb} A^a_{\wb}~, \quad T_{w\wb} =0~,}
in addition to the usual gravitational part.  In terms of  the
stress tensor we define the Virasoro generator: $L_0 -{c\over 24} =
\oint dw T_{ww}$.

Consider the shift $A^{(0)3}_w \rightarrow A^{(0)3}_w +2\eta$.
Taking $w$ to have $2\pi$ periodicity, this induces the shift
\eqn\am{\eqalign{ L_0 &\rightarrow L_0 +2 \eta J^3_0 +k\eta^2 \cr
J^3_0 & \rightarrow J^3_0 +k\eta~.}}
This is a so-called ``spectral flow" transformation.  This is an
automorphism of the usual Virasoro/current algebra, and provides yet
another justification for the boundary term \ag.

We remarked earlier that in the case of an $SO(4)\approx SU(2)\times
SU(2)$ gauge group we only considered one of the $SU(2)$ factors.
The other factor is included as follows. We add to the action \aa\ a
second Chern-Simons term with opposite sign coefficient.  The
boundary term analogous to \ag\ then implies that the current is
purely anti-holomorphic. The explicit formulas are then essentially
identical to the above, with the replacement $w\leftrightarrow \wb$.

\newsec{Higher dimensional perspective: generalities}

We now turn to the higher dimensional analysis of conserved charges, and
discuss some aspects of the problem common to the various cases.
Some previous, but not directly related,  work on a higher dimensional approach to holographic
renormalization is \refs{\TaylorRobinsonFE,\SkenderisUY,\SkenderisDI}.

We first need to discuss the class of spacetimes we will be considering.
Since charges in gauge theories are expressed as surface integrals,
what matters to us is the asymptotic behavior of the metric and matter
fields.  Our first assumption is that the metric is asymptotically, locally,
AdS$_q\times S^p$, by which we mean
\eqn\ya{ ds^2 \rightarrow ds_{AdS_q}^2 + \ell_p^2 (dy^i-A^{ij}(x)y^j)(dy^i-A^{ik}(x)y^k)~.}
AdS coordinates are denoted by $x$. Sphere coordinates are denoted
as  $y^i$; $i = 1 \cdots p + 1$; $\sum y^i y^i = 1$. We should in
principle specify the rate of  falloff of fluctuations around this
form, but this will not be necessary.

$SO(p+1)$ acts on $y^i$ in the obvious way.    We identify
$A^{ij}(x)=-A^{ji}(x)$ as Kaluza-Klein $SO(p+1)$ gauge fields by
noting that under an $x$-dependent $SO(p+1)$ rotation of $y^i$,
invariance of the line element \ya\ is achieved by accompanying
this with the usual $SO(p+1)$ gauge transformation of $A^{ij}(x)$.

We also need to specify the asymptotic form  of the field strength
$G_p$; this is a good deal more subtle and is the main topic of
the remainder of the paper.

Conserved charges  arise as integrals of conserved currents, which
are in turn defined to be conjugate to the gauge potentials
$A^{ij}$.   Specifically, the on-shell variation of the action
with respect to $A^{ij}$ takes the form of a boundary integral,
which we can write as
\eqn\yc{ \delta S = {i \over 4\pi} \int_{\p AdS}\!d^{q-1}x \sqrt{g}
~J^{ij\alpha} \delta A^{ij}_\alpha~,}
where the boundary metric is $g^{(0)}_{\alpha\beta}$ as in \ab,
but we suppress the $(0)$ superscript.       Invariance of the
action under $SO(p+1)$ gauge transformations of $A^{ij}$ implies
covariant conservation of the current.  We can then define charges
by integrating the time component of the current over a spacelike
hypersurface in the usual fashion. However, we should emphasize
that in general the action need not be gauge invariant ---
variation by boundary terms is allowed --- which leads to
anomalous conservation laws.  We will see this explicitly in the
examples below.

When we specialize to the main case of interest, AdS$_3 \times
S^p$ (with $p=2,3$), we  make the  further assumption  that the
$A^{ij}(x)$ appearing in \ya\ are flat connections; that is, the
associated field strength vanishes.   This is justified as
follows.  For reasons that will become clear as we proceed,
$A^{ij}$ asymptotically obeys a Chern-Simons equation of motion,
and this imposes flatness. Our analysis in the AdS$_3 \times S^p$
case will then proceed in parallel to that of the previous
section.    After adding a boundary term analogous to \ag, the
currents will take forms similar to \ah-\ai, and we can define
charges as before.

The main subtlety in arriving at the correct variation \yc\ lies in determining how
the $G_p$ dependent part of the action varies.  We now describe  the tools
used in this analysis.

\newsec{Review of global angular forms}

In this section we introduce the global angular form and establish
conventions and notation for dealing with sphere bundles. Our
discussion  follows  \refs{\FreedTG,\HarveyBX} but with different
normalization conventions.

We will be concerned here with $AdS_q \times S^p$ as an $S^p$ bundle
over the base space $AdS_q$ together with a connection one-form $A$
taking values in the Lie Algebra $so(p + 1)$.  Sphere coordinates
are denoted as  $y^i$; $i = 1 \cdots p + 1$; $\sum y^i y^i = 1$.
 $SO(p + 1)$ acts in the obvious  way.

The connection $A$ allows us to define vertical forms along the
$S^p$ and a curvature for the connection:

\eqn\ba{\eqalign{ Dy^i &= dy^i - A^{ij}y^j \cr F^{ij} &= [D,D]^{ij}
= dA^{ij} - A^{ik} \wedge A^{kj}~. }}
The connection $A$ is most easily understood as representing the off
diagonal components of the metric, as in \ya.
We use $x$ to denote coordinates on the AdS base. Then  $A^{ij} =
A^{ij} (x)$ is a function of the AdS coordinates only.

We are  most interested in the $SO(p+1)$
transformations that are implemented by a combination of a gauge
transformation and a coordinate transformation. Explicitly, given an
antisymmetric matrix in $so(p+1)$, $\Lambda^{ij}(x)$,we perform:
\eqn\baa{\eqalign{ y^i & \rightarrow y^i + \Lambda^{ij}y^j~, \cr
A^{ij} &\rightarrow A^{ij} + d \Lambda^{ij} + [\Lambda, A]^{ij}~. }
}
$Dy^i$ and $F^{ij}$ of course transform covariantly under \baa.

Over any oriented $S^p$-bundle it is possible to uniquely define a global
angular $p$-form, $e_p$, such that:

$\bullet$ The integral of $e_p$ over any fiber is given by
$\int_{S^p} e_p = 1$~.

$\bullet$  $d e_{2n} = 0$~.

$\bullet$  $d e_{2n -1} = \chi_{2n}$, where $\chi_{2n}$ is the Euler
class of the sphere bundle.

$\bullet$  $e_p$ is invariant under \baa.

These properties make $e_p$ well suited for writing an ${\it
ansatz}$ for the p-form field strength $G_p$ supporting an
AdS$_q\times S^p$ solution of supergravity
\refs{\HarveyBX,\NastaseCB,\CveticDM}.

Our main examples will concern the cases $p = 2,3$, for which
\eqn\bb{\eqalign{ e_2 & = {1 \over 8\pi} \epsilon_{ijk} \left(
Dy^i Dy^j- F^{ij}\right)y^k \cr de_2 & =0 \cr  e_3 &  = {1 \over
(2\pi)^2 } \epsilon_{ijkl} \left( {1 \over 3} Dy^i Dy^j Dy^k-{1
\over 2} F^{ij} Dy^k \right)y^l \cr de_3 & = \chi_4 = {1 \over
32\pi^2} \epsilon_{ijkl} F^{ij} F^{kl}~. }}
It will also be useful to have an explicit expression for $\chi_3$,
defined by $d\chi_3 =\chi_4$.  This is most naturally expressed in
$SU(2)_L\times SU(2)_R$ notatation, as defined in the appendix.  We
then have, up to a closed form,
\eqn\bbz{\chi_3  = -{1 \over
 8\pi^2}\Tr(A_L dA_L +{2 \over
3}A_L^3) +{1 \over
 8\pi^2}\Tr(A_R dA_R +{2 \over 3}A_R^3) ~.}

\subsec{ Bott and Catteneo formula }

We now state a formula due to Bott and Catteneo \Bott\ that will prove very
useful in the case of even dimensional spheres, $p=2n$.
We may write, at the level of forms:
\eqn\am{ \int_{S^{2n}} e_{2n}\wedge e_{2n} \wedge e_{2n} = {1 \over 4} p_n}
where $p_n$ is the Pontrjagin class of the sphere bundle.

We now apply ``anomaly descent" to both sides of this formula.
Given an invariant closed form like $e_{2n}$, locally  we can
write $e_{2n} = d e_{2n -1}^{(0)}$.  The invariance of $e_{2n}$
under \baa\ implies that $\delta e_{2n -1}^{(0)} = d e_{2n
-2}^{(1)}$.  We proceed in analogous fashion  for $p_n^{(0)}$.
Then we can write
\eqn\an{ \int_{S^{2n}}  e^{(0)}_{2n-1} \wedge e_{2n}\wedge e_{2n}
= {1 \over 4}p^{(0)}_n}
up to a closed form. Note that we are not relabeling the $n$
subscript on $p$ by convention; $p_n^{(0)}$ is a $4n-1$ form.

In the $n=1$ case it is convenient to work in $SU(2)$ language by writing
\eqn\ao{ A^a =\half \epsilon^{abc}A^{bc}~,\quad A = A^a {i\over 2}\sigma^a~.}
The $n=1$ version of \an\ is then
\eqn\aoa{ \int \! e_1^{(0)} \wedge e_2 \wedge e_2 = -{1 \over
2}\left({1 \over 2\pi}\right)^2 \int \Tr(A dA +{2 \over 3} A^3)~.}
If we now equate the gauge variations of both sides of this
equation we get:
\eqn\aob{\int  e^{(1)}_{0}\wedge e_{2}\wedge e_{2} = -{1 \over
2}{\left({1 \over 2\pi}\right)^2}\int \Tr(\Lambda dA)~.}

The $n=2$ Bott-Catteneo formula was used in \HarveyBX\ to derive the
Chern-Simons terms for AdS$_7 \times S^4$ spacetimes. By a similar
procedure, we  will  use the formula  for deriving the conserved
charges associated with AdS$_3\times S^2$ spacetimes, or more
generally for AdS$_{4n-1}\times S^{2n}$.

\newsec{ Example: AdS$_3 \times S^{2}$}

In this section we show how to derive the $SO(3)$ charges
associated with asymptotically, locally,
 AdS$_3 \times S^{2}$ geometries.   These geometries are important in
 string theory since they describe the near horizon limit of four dimensional black holes
 (e.g. \BalasubramanianEE).
We will also discuss the generalization to AdS$_{4n-1} \times
S^{2n}$.

\subsec{$AdS^3 \times S^2$}

We begin with the action
\eqn\ca{ S = {1 \over 16 \pi G_5} \int \!d^5x\left(\sqrt{g}R+\half
\star G_2 \wedge G_2 +i\alpha C_1 \wedge G_2 \wedge G_2 \right)~. }
As was noted earlier, the Chern-Simons term is crucial for
obtaining nonzero conserved charges induced by diffeomorphisms on
the sphere.  We leave its coefficient $\alpha$ unspecified, although in specific
constructions it is fixed by supersymmetry.

The metric takes the asymptotic form \ya\ with $q=3$ and $p=2$, and we now discuss the
asymptotic form of the field strength  $G_2$.      For $A^{ij}=0$ we have
the ``background" solution with
\eqn\caa{ G_2 = Q\epsilon_{S^2}~,}
where $\epsilon_{S^2}$ denotes the volume form on the unit
2-sphere.  The question is how to modify this in the presence of
nonzero $A^{ij}$.  Here we can follow \refs{\FreedTG,\HarveyBX}.
We want
the action for $A^{ij}$ to be invariant, up to boundary terms,
under $\delta_\Lambda A = d\Lambda +[\Lambda,A]$, in order to
define a conserved current (or rather, a current that is
anomalously conserved in the presence of a nonzero boundary
variation).  Now, since our action is diffeomorphism invariant,
this invariance will be achieved provided that the action is
invariant under \baa.\foot{We will now refer to \baa\ as a gauge
transformation.} This in turn suggests that we should demand that
$G_2$ be gauge invariant.   Furthermore, in order to construct
solutions of a fixed charge, we demand that $\int_{S^2} G_2 = 4\pi Q
$, in accordance with \caa. Finally, we must of course have
$dG_2=0$.

These conditions lead us uniquely to:
\eqn\cb{ G_2 = 4 \pi Q e_2~, }
with $e_2$ defined in \bb.    We emphasize that we are only
demanding that $G_2$ take this form asymptotically; deep in the
interior $G_2$ will generally deviate from this.

Although $G_2$ is gauge invariant, $C_1$ is not.  Indeed, we have
$C_1 = 4\pi Q e_1^{(0)}$ and

\eqn\cd{
\delta_{\Lambda} C_1 = 4 \pi Q d e_0^{(1)} ~,
}
and so the action varies by a boundary term
\eqn\ce{\delta_\Lambda S = -{i\alpha Q^3 \over 2G_5}\int_{\p AdS}\!
\Tr  (\Lambda dA) ~,}
where we used \aob.       We can use this variation to fix the coefficient of the
Chern-Simons term in the effective three dimensional action:
\eqn\cea{S_{CS} =  -{ik \over 4\pi}\int_{ AdS}\! d^3x~ \Tr(AdA+{2
\over 3}A^3)~.}
with
\eqn\ceb{ k = { 2\pi \alpha Q^3 \over G_5}~.}
In the three dimensional analysis of section 2, the currents were
obtained from the Chern-Simons term.   To identify the current and
charges in the 5-dimensional setup we can now simply follow the
analysis in section 2.

Alternatively, we can obtain the current directly by examining the
on-shell variation of the action under an arbitrary variation  of $A$.
Given the form of our ansatz for $G_2$, the only term in the
action which contributes is the Chern-Simons term. Since the
Einstein-Hilbert and $G_2$ kinetic terms are gauge invariant,
their variations are proportional to field strengths, and we have
already noted that these vanish at the boundary. The variation of
the Chern-Simons term can be evaluated  using the formula of Bott
and Cattaneo:
\eqn\ce{\eqalign{\delta S= \delta \left({ i \alpha \over 16 \pi
G_5 } \int C_1 \wedge G_2 \wedge G_2 \right)  &=  {4i \alpha \pi^2
Q^3 \over G_5} \delta \int e_1^{(0)} \wedge e_2 \wedge e_2 \cr &=
{ik \over 4\pi} \int_{\p AdS} \Tr(A \delta A )~.}}
Note that this is the same formula
as obtained by varying \cea.

We can now proceed precisely as in section 2.  After adding the
boundary term \ag\ and going to conformal gauge, we obtain the
current \ai, and the $SU(2)$ charges \ak.  We have therefore
succeeded in finding formulas for the $SO(3)$ charges with the
desired properties.   In particular, since flat potentials
$A^{ij}$ can yield nonzero charges, we see how charges can be
induced by coordinate transformations.  An explicit example of
this  will be given below in the AdS$_3\times S^3$ context.

\subsec{Generalization to AdS$_{4n-1}\times S^{2n}$}

The preceding analysis admits a straightforward generalization.  We take
the action
\eqn\caa{ S = {1 \over 16 \pi G_{6n-1}} \int \!d^{6n-1}x\left(\sqrt{g}R+\half
\star G_{2n} \wedge G_{2n} +i\alpha C_{2n-1} \wedge G_{2n}\wedge G_{2n} \right)~. }
Proceeding as above, we are led to
\eqn\cab{ G_{2n} = \Omega_{2n} e_{2n}~.}
Application of the Bott-Cattaneo formula leads to the Chern-Simons term
\eqn\cac{S_{CS} =  {i \alpha (\Omega_{2n})^3 \over 64\pi G_{6n-1}} \int_{AdS} \! d^{4n-1}x
\, p_n^{(0)}~.}
Varying this with respect to $A^{ij}$ yields the currents and charges.
We resist writing the  resulting formulas as they are not particularly
illuminating.

\subsec{Comments}

In the preceding it is manifest that the existence of Chern-Simons terms in
the AdS theory is directly tied to the presence of such terms in the higher
dimensional theory.   Note that we  only considered the two-derivative
Chern-Simons terms in the higher dimensional theory, but in string/M-theory
there can be additional Chern-Simons terms with more derivatives.  These
are exactly known in many contexts, since they are connected with anomalies,
and thus can be used to compute string/quantum corrections to the
AdS Chern-Simons terms.    This is explained in \refs{\KrausZM,\KrausVZ},
where these results are used to give a simple derivation of higher derivative
corrections to black hole entropy.

\newsec{ Example: AdS$_3 \times S^{3}$}

In this section we derive an expression for the $SO(4)$ charges of
asymptotically AdS$_3\times S^3$ geometries.     The analysis
consists of a careful treatment of the 3-form field strength.  We
find it necessary to add a gauge dependent term to the naive
expression for $G_3$, and then show how this term gives rise to
nonzero conserved charges.

The action is
\eqn\va{ S= {1 \over 16\pi G_6}\int\! d^6 x \left(\sqrt{g}R + \half \star G_3\wedge G_3\right)~,}
and the background AdS$_3\times S^3$ solution is
\eqn\vb{\eqalign{ ds^2 &= ds_{AdS_3}^2 + ds_{S^3}^2 \cr G_3 &=
Q(\epsilon_{S^3} + i\star^6 \epsilon_{S^3} ) }}
where $\epsilon_{S^3}$ is the volume form on the unit 3-sphere.
The factor of $i$ comes from working in Euclidean signature.

\subsec{{\it Ansatz} for asymptotic form of $G_3$}

We now assume that the asymptotic  metric takes the form \ya, and
seek an expression for the asymptotic form of $G_3$.    As in the
previous section, we start by demanding that $G_3$ be  closed, be
gauge invariant, have a fixed integral over the $S^3$ fiber, and
reduces to \vb\ when $A^{ij}=0$.    Our first guess is therefore
\eqn\vc{ G_3 = Q \left(2 \pi^2 e_3 +  i\star^6 \eps_{S^3} \right)
+dC_{AdS}~,\quad ({\rm first~guess})~.}
The volume form on $S^3$ is defined as $\epsilon_{S^3} = {1\over
3!}\epsilon_{ijkl}Dy^i Dy^j Dy^k y^l$. The contribution $
dC_{AdS}$ representing fluctuations of  the AdS part will play no
role in our discussion, and will be suppressed henceforth.

This expression indeed satisfies the conditions stated above.  But
it suffers from an important flaw.    We require an expression not
just for $G_3$, but also for its potential $C_2$.   A glance at
the explicit expression \bb\ for $e_3$ shows that it contains
terms cubic in $A^{ij}$ with no derivatives.  This makes it
clear that if we try to write $G_3 = dC_2$ we will be forced to
write a nonlocal expression for $C_2$.    This nonlocality is
troublesome when we recall that branes will couple directly to
$C_2$ and hence be described by nonlocal actions.  While we might
perhaps be able to make sense of this, it seems preferable to seek
a modification of
 \vc\ compatible with a local expression for $C_2$.

   The root of the problem is that the closure of
$e_3$ is in some sense an accident. In general, the global angular
form $e_3$ is defined so  that $d e_3 = \chi_4$, the Euler class
of the sphere bundle.  If the dimension of the base $AdS$ is less
than 4, $\chi_4$ trivially vanishes, since $\chi_4$ is defined as
a 4-form on the base space alone.  In higher dimensional $AdS$
spaces we are forced by the closure of $G$ to write
\eqn\vd{ G_3 = Q \left(2 \pi^2 (e_3 - \chi_3 ) +  i\star^6 \eps_{S^3}
\right)~,}
where we have taken advantage of the closure of $\chi_4$ to write
$\chi_4 = d \chi_3$.  This expression is closed in any dimension
and allows for the construction of a local ansatz for $C_2$ as we
will see below.    We will therefore take \vd\ as our asymptotic
form for $G_3$.

On the other hand, recall that the original motivation for writing \vc\ was based on
the gauge invariance of $G_3$, yet in \vd\ we have just added
a term to our ansatz which is  gauge dependent.   This will
not  be a problem provided that the gauge variation of the action is
a pure boundary term, since then the equations of motion will still be gauge
invariant.    We now show that this is indeed the case.

We write the gauge variation of $\chi_3$ as $\delta_\Lambda \chi_3 = d\chi_2$,
so that
\eqn\ve{ \delta_\Lambda G_3 = -2\pi^2 Q d\chi_2~.}
The variation of the action is then
\eqn\vf{\delta_\Lambda S = {\pi Q \over 8 G_6} \int_{\p} \star G_3 \wedge \chi_2 -
 {\pi Q \over 8 G_6} \int d\star G_3 \wedge \chi_2~.  }
The second term vanishes\foot{To see this,  note that $\chi_2$ has
both legs along the AdS, and is constant on the sphere. Therefore,
this term contains a factor of $\int_{S^3} \!d \star G_3$. Now
decompose the exterior derivative as: $d=d_{S^3} +d_{AdS}$. Then,
since ${\star G_3}$ is globally defined, we have $\int_{S^3}
\!d_{S^3}{\star G}=0$.  Finally, we need
$d_{AdS}\int_{S^3}\!{\star G} = 0$.  But the part of $G_3$ with
all 3 AdS legs is $\star\eps_{S^3}$  (using that $\chi_3$ vanishes on the boundary
for solutions obeying \ac\ and $A_\eta=0$), so inside the integral we can
take ${\star G_3}= \eps_{S^3}$. The integral of the volume form
has no AdS dependence, so it is annihilated by $d_{AdS}$.},
leaving just the following boundary term:
\eqn\lc{ \delta_\Lambda S =  { \pi Q \over 8 G_6} \int_\p \star G_3\wedge \chi_2
=  {i \pi^3 Q^2 \over 4 G_6} \int_{\p AdS} \chi_2~.}
This is of course simply the gauge variation of a three
dimensional Chern-Simons term
\eqn\ld{\eqalign{ S_{CS} &=   { i\pi^3 Q^2 \over 4 G_6} \int_{
AdS} \chi_3\cr &= -{ik\over 4\pi} \int_{AdS}\! \Tr(A_L dA_L+{2
\over 3}A_L^3) +{ik\over 4\pi} \int_{AdS}\! \Tr(A_R dA_R+{2 \over
3}A_R^3)~, }}
where we used \bbz, and defined
\eqn\lda{ k = {\pi^2 Q^2 \over 8G_6}~.}

Our formula \vd\ for $G_3$ may seem a bit surprising, but we have
shown that its variation is consistent with that of a   Chern-Simons term
in the three dimensional action.  By
contrast, the naive version \vc\ has vanishing gauge variation.
In the charge analysis that follows, it will be clear that only
the modified version \vd\ will  give the desired results. It is
also worth noting that our approach generalizes quite easily to
the case of $AdS_{2n-1} \times S^{2n-1}$.    However, we should note
that since we are working in a Lagrangian formalism we cannot immediately
include examples with self-dual field strength, such as AdS$_5 \times S^5$ in IIB
supergravity.   To cover these cases we should instead work with the equations of motion;
we hope to return to this in the future.

\subsec{Comment on chiral anomalies}

We can now give an illuminating higher dimensional interpretation of
chiral anomalies in this context.  In \lc\ we showed that the action
is not invariant under gauge transformations that extend to the boundary;
this is the anomaly.   The explanation of this is that from the six-dimensional
point of view it is clear that these are not gauge transformations, since they
shift $G_3$ according to \ve.   We further observe that $\delta_\Lambda S=0$  when
$\chi_2$ is an exact form, which is when $\delta_\Lambda G_3=0$.   So the true
(non-anomalous) gauge symmetries of the three-dimensional theory are just
those that are manifest gauge symmetries of the six-dimensional theory.

\subsec{Variation of the action}

We now derive the current by computing the on-shell variation of
the action with respect to $A^{ij}$.  The variation of the
Einstein-Hilbert term in \va\ is proportional to the field
strength of $A^{ij}$; there is no contribution to the current
since the potentials are flat at the boundary.  Thus we need only
consider
\eqn\lz{\delta S = -{1 \over 16\pi G_6} \int_{\p} \star G_3 \wedge
\delta C_2~.}

To proceed we need an expression for $C_2$ with $G_3$ given in
\vd.  The basic formula needed for this is\foot{To verify this
formula it is helpful to use the $SO(4)$ invariance to set, say,
$y^4=1$.}
\eqn\ly{ e_3 -\chi_3 ={1 \over 3!} \epsilon_{ijkl}dy^i dy^j dy^k
y^l+ d\left[ {1 \over 4 \pi^2} \epsilon_{ijkl} A^{i j} dy^k y^l -{
1 \over 8 \pi^2} \eps_{i j k l} A^{i j} A^{k m} y^l y^m \right]~.}
 Now, using the
explicit expression for $e_3$ given in \bb, together with the fact
that $\chi_3$ vanishes at the boundary,\foot{We are assuming the
potentials are of the form \ac.}
the variation of the action has two terms
\eqn\lx{\eqalign{ \delta S &= -{iQ^2 \over 128 \pi G_6} \int_{\p}
\!
 \epsilon_{mnpq} \epsilon_{ijkl}dy^m dy^n dy^k y^r y^q y^l
 A^{pr}\delta A^{ij}\cr &~~~ -{iQ^2 \over 728 \pi G_6} \delta \int_\p \!
  \eps_{mnpq} dy^m dy^n dy^p y^q
  \eps_{i j k l} A^{i j} A^{k m} y^l y^m~.
 }}

The  second line of \lx\ is easily to seen to vanish: under the
integral we can replace $y^l y^m \rightarrow {1\over 4}
\delta^{lm}$, and then use $\epsilon_{ijkl}A^{ij}A^{kl}=0$. The
first line is straightforward, though a bit tedious, to work out.
After performing the $dy^i$ integration we find
\eqn\lw{\delta S = -{ik\over 16\pi} \int_{\p AdS}
\epsilon_{ijkl}A^{ij}\delta A^{kl}~,}
with $k$ given in \lda. Converting to $SU(2)_L\times SU(2)_R$
using the conventions in the appendix, the variation can be
written
\eqn\lu{ \delta S = -{ik \over 8\pi}\int_{\p AdS}\!\left( A_L^a
\delta A_L^a- A_R^a \delta A_R^a\right)~.}

The remainder of the analysis now essentially reduces to that of
section 2, taking into account the fact that we have two copies of
$SU(2)$ gauge fields appearing with opposite sign.  The boundary
term analogous to \ag\ is therefore
\eqn\agz{ S_{\rm bndy}^{\rm gauge} = -{k \over 16\pi} \int_{\p AdS}
\! d^2 x \sqrt{g} g^{\alpha\beta} \left(A^{a}_{L\alpha}
A^{a}_{L\beta}+A^{a}_{R\alpha} A^{a}_{R\beta}\right) ~.}
The currents are
\eqn\lt{\eqalign{ J_{Lw}^a  &={ik \over 2} A_{Lw}^a~,\quad
J_{L\wb}^a =0 \cr J_{Rw}^a  &=0~,\quad J_{R\wb}^a ={ik \over 2}
A_{R\wb}^a~.}}
The charges
\eqn\ls{ J_{L0}^a = \oint{dw \over 2\pi i} J_{Lw}^a~,\quad
J_{R0}^a = -\oint{d\wb \over 2\pi i} J_{R\wb}^a~,}
then obey the $SU(2)_L\times SU(2)_R$ algebra.

\subsec{Example of spectral flow}

We start with global AdS$_3 \times S^3$
\eqn\ni{ ds^2=(1+r^2/\ell^2)dt^2 + {dr^2 \over
1+r^2/\ell^2}+r^2d\xi^2 + \ell^2 (d\theta^2 + \sin^2 \theta
d\psi^2 + \cos^2\theta d\phi^2 )~.}
The angular coordinates are related to the $y^i$ as
\eqn\nj{ \eqalign{ y^1 &= \sin\theta \sin\psi \cr y^2 &=
\sin\theta \cos\psi  \cr y^3 &= \cos\theta \sin\phi \cr y^4 &=
\cos\theta \cos\phi~, }}
and we also define the complex AdS$_3$ boundary coordinate
\eqn\nk{ w = \xi + it/\ell~.}

We now consider  diffeomorphisms that implement a spectral flow:
\eqn\nq{ d\psi \rightarrow d\psi+ \eta_L dw+\eta_R d\wb ~,\quad
d\phi \rightarrow d\phi+ \eta_L dw-\eta_R d\wb~. }
To preserve the periodicities we require $2\eta_{L,R}\in \ZZ$.
This transformation induces the following gauge fields
\eqn\hr{ A^{12} =-\eta_L dw+\eta_R d\wb~,\quad A^{34} = -\eta_L
dw-\eta_R d\wb~,}
or equivalently,
\eqn\ht{A_L^3=2\eta_L dw~,\quad A_R^3 =2\eta_R d\wb ~.}
The charges are therefore
\eqn\hu{ J^3_{L0}= k\eta_L~,\quad J^3_{R0}= k\eta_R~.}
These are the correct charges induced by spectral flow (as in
\am.) Following the analysis of section 2, we also find that the
Virasoro charges transform as in \am.  This example therefore
provides a simple illustration of how coordinate transformations
can generate nonzero charges.

\subsec{General rotating solutions of D1-D5 system}

The D1-D5 system is the canonical example of an AdS$_3\times S^3$
geometry.  Comparing normalizations with, e.g. \MaldacenaBW, we
find
\eqn\ua{ k = N_1 N_5~.}
This agrees with level of the $SU(2)$ current algebras of the dual
CFT.

General solutions corresponding to black holes, black rings, or
otherwise, take the asymptotic form \ya. We typically choose
coordinates such that the nonzero $SU(2)_L\times SU(2)_R$ charges
are $J^3_{L,R 0} = \half J_{L,R}$, with $J_{L,R} \in \ZZ$. These
solutions therefore have
\eqn\ub{ A^3_L = {J_L \over k}dw~,\quad A^3_R = {J_R \over
k}d\wb~.}
These charges are conserved provided that (as is the case for the standard
black hole/ring solutions) $A^a_{L\wb}=A^a_{Rw} =0$ on the boundary.
If these components are nonzero then the currents are anomalous, and the charges
are not conserved.  This is completely consistent with the AdS$_3$/CFT$_2$
dictionary, in particular with the R-symmetry anomalies of the CFT.

\subsec{Generalization to AdS$_{2n-1}\times S^{2n-1}$}

The generalized version of \vd\ is
\eqn\uc{ G_{2n-1} = Q\left(\Omega_{2n-1}(e_{2n-1}-\chi_{2n-1}) +i \star^{4n-2} \epsilon_{S^{2n-1}}\right)~,}
where the Euler class is $\chi_{2n} = d\chi_{2n-1}$.   Following the same steps as led
to \ld, we find that the $2n-1$ dimensional action contains the Chern-Simons term
\eqn\uv{ S_{CS} = {i \Omega_{2n-1}^2 Q^2 \over 16 \pi G_{4n-2} } \int_{AdS} \chi_{2n-1}~.}
A contribution to the  current is obtained from the on-shell variation of \uv.
Two other contributions to the current come from the  option of adding a boundary term analogous to \agz, and from the variation of Maxwell type terms (note that above three dimensions,
the Chern-Simons term is no longer the term with the fewest derivatives.)
We will not
explore this further here.

\bigskip
\noindent {\bf Acknowledgments:} \medskip \noindent We thank Finn
Larsen, Don Marolf, and Simon Ross for discussions. The work of PK
is supported in part by NSF grant PHY-00-99590.

\appendix{A}{Translation between $SO(4)$ and $SU(2)_L \times SU(2)_R$}

Our $SO(4)$ generators are
\eqn\re{ J^{ij} = -i(y^i \p_j - y^j \p_i)~.}
We then define  self-dual and anti-self dual combinations:
\eqn\rf{\eqalign{ J_+^{ij} &= \half ( \half
\epsilon^{ijkl}J^{kl}+J^{ij}) \cr  J_-^{ij} &= \half ( \half
\epsilon^{ijkl}J^{kl}-J^{ij} )~.}}
In terms of these we define the generators $(a=1,2,3)$
\eqn\rg{ \eqalign{ J_L^a&= J_+^{a4} \cr    J_R^a &= J_-^{a4} }}
which obey the $SU(2)_L \times SU(2)_R$ algebra:
\eqn\rh{\eqalign{ [J_L^a,J_L^b]&= i\eps^{abc} J_L^c\cr
[J_R^a,J_R^b]&= i\eps^{abc} J_R^c \cr [J_L^a,J_R^b]&= 0~. } }
The $SO(4)$ gauge fields $A^{ij}$ are then related to the $SU(2)_L
\times SU(2)_R$  gauge fields $A_{L,R}^a$ via
\eqn\ri{ A_L^a J_L^a + A_R^a J_R^a =\half A^{ij} J^{ji}}
which yields
\eqn\rj{ A^{a4} = -\half (A_L^a -A_R^a)~,\quad A^{ab} =- \half
\epsilon^{abc}(A_L^c +A_R^c)~.}

Upon defining $A_{L,R} = A_{L,R}^a {i\over 2}\sigma^a$, we find that
\rj\ implies the relations
\eqn\rk{ \eqalign{ F^{ij}F^{ij} & = -2 \Tr F_L^2 - 2\Tr F_L^2  \cr
\chi_4
 &= {1 \over 32\pi^2} \epsilon^{ijkl}F^{ij}F^{kl}  = -{1 \over 8\pi^2}
 \Tr F_L^2 + {1 \over 8\pi^2}\Tr F_R^2  \cr \chi_3 & = -{1 \over
 8\pi^2}\Tr(A_L dA_L +{2 \over
3}A_L^3) +{1 \over
 8\pi^2}\Tr(A_R dA_R +{2 \over 3}A_R^3) }}
with $\chi_4 = d\chi_3$.

\listrefs

\end